\documentclass[aps,prl,reprint,superscriptaddress,nofootinbib,floatfix]{revtex4-2}

\usepackage{amsmath,amssymb,bm}
\usepackage{graphicx}
\usepackage{xcolor}
\definecolor{revisiongreen}{RGB}{0,125,70}
\usepackage{hyperref}
\usepackage[normalem]{ulem}
\definecolor{prllink}{RGB}{64,135,214}
\hypersetup{colorlinks=true,citecolor=prllink,linkcolor=prllink,urlcolor=prllink}
\graphicspath{{figures/}}

\newcommand{\Msun}{M_\odot}
\newcommand{\dd}{\mathrm{d}}
\newcommand{\CDMTriplets}{0.364}
\newcommand{\TransientTriplets}{0.499}
\newcommand{\RemnantTriplets}{1.66}
\newcommand{\TransientFourCopies}{$2.24\times10^{-3}$}
\newcommand{\RemnantFourCopies}{0.149}

\newcommand{\CDMFourFraction}{\ensuremath{4.3\times10^{-6}}}
\newcommand{\CDMFiveFraction}{\ensuremath{2.4\times10^{-8}}}
\newcommand{\TransientFourFraction}{\ensuremath{0.45\%}}
\newcommand{\TransientFiveFraction}{\ensuremath{0.40\%}}
\newcommand{\RemnantFourFraction}{\ensuremath{9.0\%}}
\newcommand{\RemnantFiveFraction}{\ensuremath{5.0\%}}
\newcommand{\StageEarlyOuter}{2.869}
\newcommand{\StageMiddleOuter}{2.905}
\newcommand{\StageLateOuter}{2.910}
\newcommand{\ConvRadial}{$+0.02\%$}
\newcommand{\ConvAngle}{$-5.0\%$}
\newcommand{\ConvImpact}{$-5.3\%$}
\newcommand{\ConvMedoid}{$-2.5\%$}
\newcommand{\ConvDelay}{$-0.003\%$}
\newcommand{\CriticalityOnset}{1.033}

\newcommand{\OneSecondOverlap}{$5.2\times10^{-4}$}
\newcommand{\TimingLow}{$4.1\times10^{-4}\,{\rm s}$}
\newcommand{\TimingHigh}{$1.7\times10^{-3}\,{\rm s}$}

\newcommand{\SurvivingPosteriorMedian}{0.57}

\newcommand{\EllipticalFiveMin}{0.17}
\newcommand{\EllipticalFiveMax}{0.75}

\begin{document}

\title{Gravitational-Wave Image Multiplicity as a Topological Probe of Dark-Matter Core Collapse}

\author{Tonghua Liu}
%\email{liutongh@yangtzeu.edu.cn}
\affiliation{School of Physics and Optoelectronic Engineering, Yangtze University, Jingzhou 434023, China}

\author{Kai Liao}
\email{liaokai@whu.edu.cn}
\affiliation{School of Physics and Technology, Wuhan University, Wuhan 430072, China}

\author{Marek Biesiada}
\email{marek.biesiada@ncbj.gov.pl}
\affiliation{National Centre for Nuclear Research, Pasteura 7, PL-02-093 Warsaw, Poland}

\author{Bing Sun}
\email{bingsun@mail.bnu.edu.cn}
\affiliation{Department of Basic Courses, Beijing University of Agriculture, Beijing 102206, China}

\begin{abstract}
Dark-matter self-interactions can drive subhalos through gravothermal core
collapse, but conventional lensing observables vary continuously with uncertain
inner profiles.  We show that gravitational-wave image multiplicity provides a
discrete alternative.  Near a macro-critical minimum, increasing subhalo
central convergence opens two caustics in sequence, converting one detectable
waveform into three and then five stationary images.  Using weighted SASHIMI
subhalos and an ET+CE selection model, we forecast almost 2 systems with at
least three copies detectable in five years if dense remnants survive,
compared with the prediction of 0.36 in 5yrs for standard CDM.  
The much more diagnostic second case is rare: the corresponding 5yrs yield with at least four detectable copies is 0.149, versus $1.55\times10^{-6}$ for CDM. According to the Poisson statistics, this gives only a 13.8\% probability of at
least one event detected in 5 years.  Thus this is a rare-event, exposure-limited test rather than a guaranteed detection: a resolved second pair would be difficult to reconcile
with the fiducial single-halo CDM population, whereas a five-year null result
would not by itself constrain remnant survival strongly. 
\end{abstract}

\maketitle

\textit{Introduction.---}
Self-interacting dark matter (SIDM) turns a halo into a thermodynamic system.
Scattering first carries heat inward and turns a cusp into a low-density core.
Once the center is nearly isothermal, the heat flow reverses: the core loses
energy, contracts, and heats up as it runs toward gravothermal collapse
~\cite{2000PhRvL..84.3760S,2002ApJ...568..475B,2023ApJ...946...47Y}.  Tidal
stripping can accelerate this evolution, whereas heating and evaporation by
the host can delay it~\cite{2020PhRvD.101f3009N,2022MNRAS.513.4845Z,
2025PhRvD.111f3001Z}.  {The lensing consequence is mass redistribution rather than mass growth: at fixed bound mass, negative heat capacity drives material into a smaller core and raises the projected convergence that enters the lens equation.}
Dense cores can therefore form, but their abundance
depends on a question that simulations have not settled: does the collapsed
state disappear quickly, or survive as a compact remnant?

Strong lensing has begun to see objects in precisely this regime.  Low-mass
dark satellites have already been detected by gravitational imaging
~\cite{2012Natur.481..341V}.  Radio imaging of JVAS B1938+666 found a perturber
containing
$(1.13\pm0.04)\times10^6\,\Msun$ within 80 pc
~\cite{2025NatAs...9.1714P}.  Its favored reconstruction has an unresolved
component smaller than 10 pc embedded in a 139 pc envelope
~\cite{2026NatAs..10..440V}.  A tidally stripped SIDM halo in deep
gravothermal collapse naturally produces this two-scale structure, while a
cold-dark-matter (CDM) interpretation requires a central black hole and an
extreme stripping history~\cite{2026arXiv260612909Z}.  The observation shows
that a dense dark perturber can be found.  One static object, however, cannot
reveal how long that state lasts.

That limitation is not cured by a more precise density reconstruction.  Arcs
and flux ratios measure the projected mass at one instant
~\cite{2002ApJ...572...25D,2021MNRAS.507.1662M,
2026arXiv260116818H,2026SciBu..71.1349Y}.  An unusually concentrated CDM halo, a black-hole-seeded
remnant, and a collapsed SIDM core can therefore look alike.  Extra-image
anomalies in electromagnetic lenses have been proposed as abundance probes of
compact and fuzzy dark matter, with detectability limits set by angular resolution
~\cite{2026PhRvD.113j3542H}.  Gravitational waves (GWs) provide a different
observable.  Each lensing image is a time-delayed copy of the same phase-coherent
chirp~\cite{2017NatCo...8.1148L}, so images that cannot be separated on the sky can still be counted.  A
compact perturber beside a highly magnified macro image can create extra
arrivals with definite delays and Morse phases
~\cite{1971PhRvD...3.3239L,1998PhRvL..80.1138N,2003ApJ...595.1039T,
2018PhRvD..98j4029D,2025PhRvL.134o1401L}.  The critical curve does not
merely magnify a small waveform perturbation; it changes the number of
solutions of the lens equation.

Previous works asked how a fixed halo profile modifies a lensed waveform
~\cite{2022PhRvD.106b3018G,2023PhRvD.108j3529T,
2026arXiv260621519A,2026arXiv260722787L}.  We ask instead when an evolving halo
makes new images.  {Our central observation is that gravothermal evolution
is continuous in density but discrete in lens topology.  This separates the
two pieces of physics: internal particle velocities set the gravothermal
heat-transfer rate, whereas the macro-lens eigenvalues set whether the
resulting profile creates additional stationary points.}  Core contraction produces a two-step sequence,
$1\rightarrow3\rightarrow5$, that cannot be removed by refitting the smooth
lens.  We derive the two pair-creation conditions, evaluate them along
fluid-calibrated SIDM histories, and carry the image counts into a
third-generation GW population.  The Supplemental Material gives the profiles,
pair-creation derivation, ellipticity and solver tests, detector selection,
population and endpoint calculation, background models, likelihood, control
observables, and convergence checks in Secs.~I--XI~\cite{supplement}.

\textit{Image-pair creation.---}
Near one macro image (one of the galaxy-scale images made by the smooth lens),
the galaxy lens supplies two eigenvalues of the local inverse-magnification
matrix,
$\lambda_t=1-\kappa_{\rm macro}-\gamma_{\rm macro}$ and
$\lambda_r=1-\kappa_{\rm macro}+\gamma_{\rm macro}$.  Here
$\kappa_{\rm macro}$ and $\gamma_{\rm macro}$ are the local convergence and
shear, and the subscripts $t$ and $r$ denote the tangential and radial
directions.  A critical curve is where one of these eigenvalues vanishes; its
mapping into the source plane is a caustic.  {Physically, $\lambda_t$ and $\lambda_r$ are the two local restoring curvatures of the smooth Fermat surface.  The subhalo mean convergence opposes those curvatures, and a zero crossing creates a new stationary-point pair.}  Let $\bm x$ be the angular
position in the image plane relative
to the subhalo, $\bm y$ the corresponding source-plane offset, and
$r=|\bm x|$.  In the principal axes of the shear, the lens equation for an
axisymmetric subhalo is
\begin{equation}
 \bm y=
 \begin{pmatrix}\lambda_t&0\\0&\lambda_r\end{pmatrix}\bm x
 -\alpha_{\rm sub}(r)\frac{\bm x}{r},
 \qquad \mu_{\rm macro}=(\lambda_t\lambda_r)^{-1},
\end{equation}
where $\alpha_{\rm sub}(r)$ is the subhalo deflection angle and
$\mu_{\rm macro}$ is the signed magnification of the smooth macro image.  The
mean subhalo convergence inside $r$ is
$\bar\kappa_{\rm sub}(<r)=\alpha_{\rm sub}(r)/r$.  Away from a
principal axis, images appear in pairs whenever the rising mean convergence
crosses one of these eigenvalues.  For a positive-parity macro image,
$0<\lambda_t<\lambda_r$.  Crossing $\lambda_t$ creates a minimum--saddle pair
and opens a three-image caustic.  Creating a second, saddle--maximum pair
requires
\begin{equation}
 \max_r\bar\kappa_{\rm sub}(<r)>\lambda_r .
 \label{eq:threshold}
\end{equation}
{Because $\lambda_r>\lambda_t$, the second crossing requires a denser core and is a more selective probe of late collapse.}  The critical curve therefore acts as a bifurcation amplifier: a gradual rise
of the central density becomes the discrete sequence
$1\rightarrow3\rightarrow5$.  For a macro saddle,
$\lambda_t<0<\lambda_r$; positive subhalo convergence cannot cross both
eigenvalues, and this sequence is absent.  Equation~\eqref{eq:threshold}
depends on the compact core relative to its local galaxy environment, not on
an assigned concentration or on an arbitrary definition of collapse time.
It is exact for the circular population model.  The derivation of the two
thresholds and the treatment of macro saddles are given in Supplemental
Material~\cite{supplement}, Sec.~II.  A two-dimensional test with
projected axis ratios down to 0.8~\cite{1997ApJ...482..604K} retains the
five-image branch in every orientation, although its source coverage changes
from \EllipticalFiveMin\ to \EllipticalFiveMax.  Supplemental Material
~\cite{supplement}, Sec.~III and Fig.~S1 give this ellipticity test.

\begin{figure*}[t]
 \includegraphics[width=\textwidth]{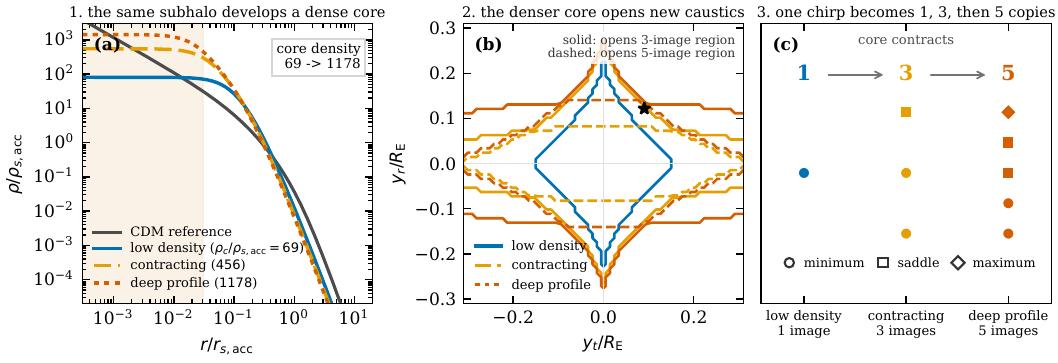}
 \caption{{Panels follow one fixed source from the evolving density profile, through its caustics, to the resulting stationary points.} A macro critical curve converts continuous core contraction into two
 discrete image-pair creation events.
 (a) Three fluid-calibrated SIDM profiles
 after outside-in tidal stripping; labels give
 $\rho_c/\rho_{s,\rm acc}$, and the gray curve is the mass-matched CDM halo.
 (b) Source-plane caustics for $\mu_{\rm macro}=3$ and $\lambda_r=1$.  Solid and
 dashed curves enclose the three- and five-image regions; the star marks one
 fixed source.  (c) The actual $1\rightarrow3\rightarrow5$ root sequence for
 that source.  Vertical placement is schematic; circles, squares, and diamonds
 give the computed Morse types: minimum, saddle, and maximum.}
 \label{fig:physics}
\end{figure*}

We project gravothermal fluid solutions calibrated to SIDM simulations
~\cite{2023ApJ...946...47Y,2025arXiv250413004M}.  Each profile is truncated
outside-in at a radius $r_t$ until it contains the subhalo's
bound mass at the lens epoch, preserving the collapsed center
while removing the stripped envelope.
The CDM control is a smoothly truncated Navarro--Frenk--White (NFW) halo of the
same bound mass~\cite{1997ApJ...490..493N,2009JCAP...01..015B}.  The analytic
profile and all parameters are given in Supplemental Material~\cite{supplement},
Sec.~I and Table~S1; the associated truncation and convergence checks are
shown in Fig.~S3.  We use $\rho_c$ and $r_c$ for the central density and
core radius, and $\rho_{s,\rm acc}$ and $r_{s,\rm acc}$ for the NFW scale
density and scale radius of the progenitor subhalo when it
entered the host halo.
Figure~\ref{fig:physics} shows the mechanism for one representative
model subhalo.
The three profiles have the same bound mass, so their outer envelopes
nearly coincide [Fig.~\ref{fig:physics}(a)].  Only the core changes: its central
density rises by a factor of 17 while its radius contracts by almost a factor
of four.  This distinction is essential.  The signal is driven by a local
thermodynamic transformation, not by adding mass to the subhalo.  {Contraction moves projected mass inward and increases $\max_r\bar\kappa_{\rm sub}(<r)$ while leaving the tidally fixed outer envelope nearly unchanged.}

Placed next to a positive-parity macro image, the contraction itself first opens
a three-image caustic and then a nested five-image region
[Fig.~\ref{fig:physics}(b)].  The star fixes the source position: the image
number changes because the core evolves, not because the alignment is tuned.

For this source, the two crossings create the required minimum--saddle and
saddle--maximum pairs [Fig.~\ref{fig:physics}(c)].  At every stage,
$N_{\rm min}+N_{\rm max}-N_{\rm sad}=1$, where the three symbols count
minimum, maximum, and saddle images, respectively.  The sequence is therefore
a change in the mapping itself: it creates additional delayed
GW copies rather than merely rescaling the strain amplitude
~\cite{2021MNRAS.506.5430J}.

\textit{From image pairs to repeated chirps.---}
For the $10^6$--$10^{10}\,\Msun$ stripped remnants considered here, the
diffraction transition lies near $10^{-7}$--$10^{-2}\,{\rm Hz}$, far below
the band probed by ground-based detectors.  ET+CE therefore sees each stationary image as a
separate arrival,
\begin{equation}
 \widetilde h_{\rm L}(f)=\widetilde h_{\rm src}(f)
 \sum_j |\mu_j|^{1/2}
 \exp\!\left(2\pi i f\Delta t_j-i\pi n_j\right),
\end{equation}
Here $f$ is the observed GW frequency.  The quantity
$\widetilde h_{\rm src}$ is the unlensed waveform, and $\mu_j$,
$\Delta t_j$, and $n_j$ are the magnification, relative delay, and Morse index
of image $j$.  The values $n_j=0,1/2,1$ identify a minimum, saddle, and
maximum.  We count an image when its network signal-to-noise ratio (SNR)
exceeds 8 and merge arrivals separated by less than one second.  Across nine
mass quantiles of the public binary-black-hole (BBH) catalog, the overlap of
copies separated by one second is below \OneSecondOverlap.  Their timing error
at SNR 8 is only \TimingLow--\TimingHigh, far below the one-second clustering
interval.
The stationary-image and detector checks are shown in Supplemental Material
~\cite{supplement}, Secs.~IV--V and Fig.~S2.

The loudest arrival can trigger a search for the same detector-frame masses,
spins, and sky position in the remaining data.

\begin{figure*}[t]
 \includegraphics[width=\textwidth]{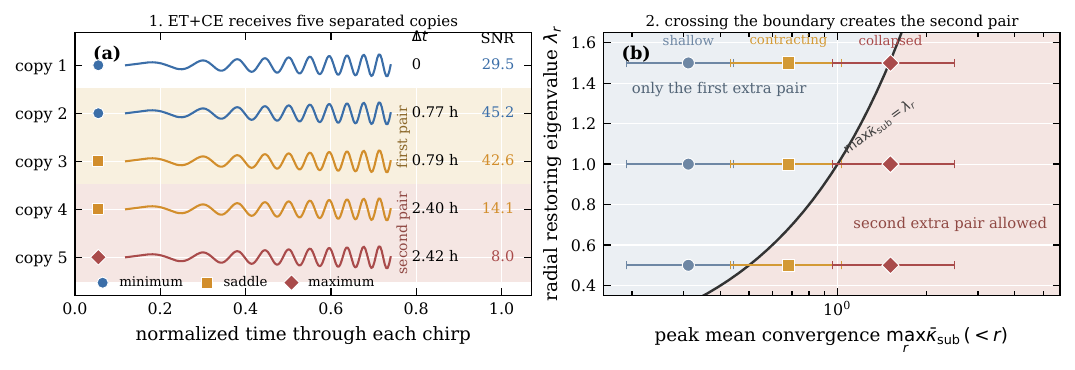}
 \caption{{Panel (a) follows one source, whereas panel (b) locates the same pair-creation threshold across the weighted population.} The detector signal and its physical threshold.  (a) Five arrivals
 of the same normalized chirp; labels give the computed delay and network SNR,
 and symbols give the Morse type.  Chirp %widths
 signal plots
 are schematic.  The two shaded
 pairs are created successively as the core contracts.  (b) Pair-creation phase
 diagram for $|\mu_{\rm macro}|=3$.  Markers show population-weighted medians at
 three fluid stages and horizontal bars contain 68\% of the weighted model subhalos.
 The second pair exists only below this boundary in the plotted coordinates,
 $\lambda_r=\max_r\bar\kappa_{\rm sub}(<r)$.}
 \label{fig:echoes}
\end{figure*}

Figure~\ref{fig:echoes}(a) shows the observable in its simplest form.  The same
intrinsic chirp appears five times.  The first new pair arrives after
$0.77\,\mathrm{h}$; the second follows near $2.4\,\mathrm{h}$.  All five copies
pass the adopted threshold, including the last maximum at SNR 8.0.  Even the
closest pair is separated by about one minute, far longer than the timing
uncertainty.  The rows are therefore separate detections, not interference
fringes within one waveform.

Panel (b) shows why the fifth image is a statement about the core rather than
an accident of alignment.  The horizontal axis is the largest mean
convergence reached by the subhalo; the vertical axis is the radial restoring
eigenvalue of the surrounding galaxy.  Each marker gives the
population-weighted median for one stage of core evolution, and each bar spans
the central 68\% of the weighted model subhalos.  Contraction moves the population
to the right.  The %diagonal
black solid line is Eq.~\eqref{eq:threshold}: only after it is
crossed can the second pair exist.  Crossing the local
macro-lens eigenvalue $\lambda_r$ therefore converts the continuous rise in
core convergence into the discrete birth of the second image
pair.

\textit{A population test of the collapsed phase.---}
The public semi-analytical  Semi-Analytical SubHalo Inference ModelIng (SASHIMI) calculation follows
subhalo accretion and tidal evolution~\cite{2022MNRAS.517.2728H}.  Each
weighted model subhalo retains its accretion time and mass,
tidal mass loss, lens-epoch bound mass, scale radius, and
tidal radius.  We retain twelve
representative model subhalos and their population weights.
Collapse
probabilities are taken from velocity-dependent SIDM simulations
~\cite{2025PhRvD.111f3001Z}; we assign no collapse to progenitors below the
simulations' $10^8\,\Msun$ accretion-mass limit.  {The velocity entering the internal heat-transfer cross section is the relative speed of SIDM particles inside a subhalo, not the subhalo's bulk orbital speed through its host; the latter controls host--subhalo evaporation and is already included in the simulations supplying these collapse probabilities.}

The GW sample we use is the public ten-year ET+CE strong-lens catalog of
Ref.~\cite{2026ApJS..284...56L}.  For each binary-black-hole (BBH) macro image,
we combine its SNR, parity, magnification, and local eigenvalues with the
weighted model subhalos.  We integrate the source-plane area that produces
at least three, four, or five detectable copies and sum the splitting
probability over the catalog.  A halo selected to collapse is first removed
from the CDM population before its evolved SIDM profile is inserted.  For
$m=3,4,5$, the resulting $N_{\geq m}$ are the expected counts of BBH
macro-image candidates with at least $m$ detected copies; their local lens
environments can differ even within the same lens system.  Supplemental
Material~\cite{supplement}, Sec.~VI, gives the full probability and source-plane
integral; endpoint weighting and five-year rates are given in Sec.~VII and
Table~S2.

The unresolved endpoint can be stated without assigning a single lifetime {to every collapsed halo}.  Let $f_{\rm surv}$ be the fraction of collapse-selected
halos whose compact endpoint is still present when they are lensed.  The
expected counts then interpolate between the collapsed-phase average and the
long-lived endpoint,
\begin{equation}
 N_{\geq m}(f_{\rm surv})=N_{\geq m}^{\rm tr}
 +f_{\rm surv}\bigl(N_{\geq m}^{\rm rem}-N_{\geq m}^{\rm tr}\bigr).
 \label{eq:survival}
\end{equation}
Here ``transient'' weights every collapsed profile by the time spent there
after the simulation's collapse threshold; ``remnant'' lets the deepest
calibrated profile survive.  Equation~\eqref{eq:survival} describes a population
mixture, not an interpolation between density profiles.

\begin{figure*}[t]
 \includegraphics[width=\textwidth]{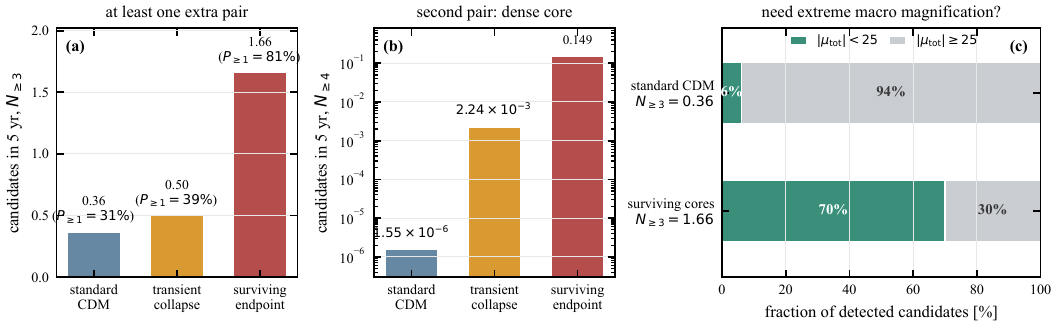}
 \caption{The result regarding three directly observable settings.   (a) Five-year
 yield of macro-image candidates with at least three detected copies; labels
 give the probability of at least one candidate.  (b) {Five-year yield
 with at least four copies, which requires the second image pair; the
 logarithmic axis allows to illustrate the orders-of-magnitude differences as displayed by numbers above the bars.}  (c) {Fraction of detected
 candidates below and above $|\mu_{\rm tot}|=25$; row labels retain the
 absolute $N_{\geq3}$.}  Ordinary CDM needs an extreme macro critical field,
 whereas a collapsed core does not.}
 \label{fig:survival}
\end{figure*}

The two pair births play different observational roles.  The first pair
supplies the triplet sample; the second identifies the densest members of that
sample.
{Figure~\ref{fig:survival}(a) first shows the absolute number of candidates
with at least one additional image pair.}  In five years,
$N_{\geq3}$ rises from \TransientTriplets\ to \RemnantTriplets\ as
$f_{\rm surv}$ goes from zero to one.  The corresponding chance of at least
one candidate rises from 39\% to 81\%.

{Panel (b) then isolates the second pair, the more selective signature of
a dense collapsed core.}  The fraction of candidates retaining at least four detectable copies rises from
\TransientFourFraction\ for a transient dense phase to \RemnantFourFraction\
for a surviving endpoint; standard CDM gives only \CDMFourFraction.  This is
the second-pair measurement.  {In absolute terms, $N_{\geq4}$ rises from
$1.55\times10^{-6}$ for standard CDM to \TransientFourCopies\ for a transient
phase and \RemnantFourCopies\ for a surviving endpoint.}

Five stationary images need not give five detected copies: the faintest can
fall below SNR 8 or merge with a neighbor.  We therefore use the disjoint
populations with exactly three and at least four detected copies, and
marginalize the rate calibration and CDM concentration.  Supplemental
Material~\cite{supplement}, Sec.~IX and Fig.~S4 give the full likelihood.

The CDM comparison uses the same model subhalos, macro images, source
positions, SNR cut, and time-delay cut.  The cosmological concentration--mass
relation~\cite{2019ApJ...871..168D} gives
$(N_{\geq3},N_{\geq4})=(\CDMTriplets,1.55\times10^{-6})$.  Artificially
rescaled concentrations are retained only as stress tests in Supplemental
Material~\cite{supplement}, Sec.~VIII and Table~S3; they are not CDM
population models.

Figure~\ref{fig:survival}(b) exposes a second physical difference.  Let us define
$|\mu_{\rm tot}|$ as the sum of the absolute magnifications of the smooth macro
images in one lens system.  {Panel (c) asks the direct observational
question: do the candidates require an extreme macro\sout{-critical} environment?}
Standard CDM produces almost all of its candidates
in extremely magnified systems:
only 0.022 are expected below $|\mu_{\rm tot}|=25$ in five years.  A surviving
collapsed-core population gives 1.16 candidates below that threshold.  Thus
70\% of its candidates lie in the lower-magnification
region, but
only 6\% of standard-CDM candidates obey this criterion.

This split also addresses compound and line-of-sight structure.  Published
compound-CDM catastrophes are selected at $|\mu_{\rm tot}|>25$
~\cite{2025arXiv251014953V}.  We additionally place CDM field halos in 100
independent light cones, following the standard line-of-sight halo treatment
~\cite{2019MNRAS.487.5721G}, and solve the full multi-plane lens equation for
eight lower-magnification sources.  The 800 solutions produce no additional bright
pair localized on a field halo; the 95\% binomial upper limit is 0.37\% per
tested configuration.  Line-of-sight shear can change the number of macro
images, but those additional images are not centered on a field halo and
belong in the macro model, not the unresolved local-copy sample.  Supplemental
Material~\cite{supplement}, Sec.~VIII, Table~S4, and Sec.~X with Fig.~S5 give
the calculation and its limitations.

JVAS B1938+666 exposes another degeneracy: its two-scale profile can arise
from collapsed SIDM or an extremely stripped, black-hole-seeded CDM remnant
~\cite{2026NatAs..10..440V,2026arXiv260612909Z}.  Image topology depends on
total convergence, so these remnants must enter the population analysis.

One high-multiplicity candidate would establish a compact perturber, not identify
SIDM by itself.  The dark-matter test comes from the population: the candidate
rate, the fraction that cross the second threshold, macro-image parity, and
the distribution of macro magnification.  Independent information on halo
concentration remains essential.

\textit{Discussion.---}
{The central result is that gravothermal collapse is continuous in density
but discrete in lens topology.}
The new observable is not another estimate of the inner density slope.  It is
the fraction of a population that crosses successive lensing bifurcations.
Slowly varying
wave-optics differences between NFW and SIDM halos may be absorbed when the
smooth lens is refitted~\cite{2026arXiv260722787L}; a newly born image pair
cannot.  Copy number and {Morse type} turn continuous gravothermal evolution
into a discrete detector signal.

Electromagnetic and GW lenses then answer different questions.  Resolved arcs
can establish that one compact object exists.  Repeated chirps reveal how often
the population crosses each pair-creation boundary.  The ratio
$N_{\ge4}/N_{\ge3}$ {is preferentially sensitive to} the occupancy of the deepest collapsed state.

For the masses and frequencies considered here, the diffraction regime criterion places ET+CE in the geometric-optics regime; the computed delays also exceed the adopted arrival-clustering interval.  The observable is therefore a set of delayed stationary-image chirps, not a single coherent interference pattern.
Ground-based detectors receive separate copies minutes to hours apart and
recognize them through the same intrinsic chirp.

The measurement is rate limited and not background free.  Subthreshold
recovery increases the matched yield by at most 16\% in our tests and cannot
rescue it.  A surviving endpoint still gives only a 13.8\% chance of at least
one four-copy candidate in five years, and the
inference must include baryonic satellites, line-of-sight halos, compound CDM
caustics, black-hole-seeded remnants, and the macro lens.  These complications
do not erase the basic test.  A null sample {constrains the lens-epoch
occupancy of compact collapsed cores under the stated population model}; a
population that follows the two pair births {tests that occupancy through
both topology and environment}.  Numerical convergence and the remaining population checks are
collected in Supplemental Material~\cite{supplement}, Sec.~XI and Table~S5.
Image topology therefore makes a long-lived collapsed endpoint
falsifiable at the population level, while the absolute rate remains observing cadence limited.

\textit{Acknowledgments.---}
This work was supported by the National Natural Science Foundation of China
(Grant No. 12675061), the National Key R\&D Program of China
(No. 2024YFC2207400), and the Polish National Science Centre
(Grant No. 2023/50/A/ST9/00579).

\textit{Data availability.---}
The figure data and code are available from the author during review and will
be deposited publicly before publication.

\nocite{2020A&A...641A...6P,2020PhRvD.102f4001P}
\bibliography{references}

\appendix
\section*{Supplemental Material}
\setcounter{equation}{0}\setcounter{figure}{0}\setcounter{table}{0}
\renewcommand{\theequation}{S\arabic{equation}}
\renewcommand{\thefigure}{S\arabic{figure}}
\renewcommand{\thetable}{S\arabic{table}}

This Supplemental Material gives the density profiles, stationary-image
solver, detector selection, and event-rate calculation used in the Letter.  It
also separates inputs taken from published {models and} simulations from the two endpoint
assumptions introduced here.

\section{Density profiles and tidal truncation}

Each {weighted subhalo model generated by the public
Semi-Analytical SubHalo Inference ModelIng (SASHIMI) calculation}~\cite{2022MNRAS.517.2728H} carries the accretion mass $M_{\rm acc}$, accretion
redshift, accretion scale radius $r_{s,\rm acc}$, and accretion scale density
$\rho_{s,\rm acc}$, together with the {bound mass at the lens
epoch} $M_{\rm sub}$, CDM scale radius $r_s$, scale density $\rho_s$, and tidal radius.  We keep these
properties attached to {each particular {model subhalo}}; a median mass is never combined with
the scale radius or stripping factor of another subhalo.

Here $r$ refers to the radius in {three dimensions}.  The quantity $M_{\rm acc}$ is the
{progenitor mass when the subhalo first enters its host halo},
$M_{\rm sub}$ is its {bound mass at the lens epoch}, $r_s$ and $\rho_s$ are the
{lens-epoch NFW scale radius and density}, and $r_t$ is the radius at which the profile is
tidally truncated.  The quantities {$r_{s,\rm acc}$ and
$\rho_{s,\rm acc}$ are the progenitor subhalo's NFW scale radius and scale
density when it enters the host halo}.  In the gravothermal profiles, $\rho_c$ and $r_c$ denote
the central density and core radius, while $r_{\rm out}$ is the outer radius
of the displayed profile.

The CDM control is the smoothly truncated NFW profile
~\cite{1997ApJ...490..493N,2009JCAP...01..015B},
\begin{equation}
 \rho_{\rm CDM}(r)=
 \frac{\rho_s}{(r/r_s)(1+r/r_s)^2}
 \frac{r_t^2}{r^2+r_t^2} .
 \label{eq:tnfw}
\end{equation}
We choose the smooth truncation radius so that Eq.~\eqref{eq:tnfw} has the
same bound mass as the sharply truncated SASHIMI {model subhalo}.

The SIDM profiles follow the late gravothermal solution of
Refs.~\cite{2023ApJ...946...47Y,2025arXiv250413004M}.  In the notation of that
solution, $\widehat\sigma$ and $\widehat t$ are the dimensionless scattering
cross section and time, while $\beta$ calibrates heat conduction in the
long-mean-free-path regime.  Thus $\beta\widehat\sigma\widehat t$ is a
dimensionless ordering variable for the fluid sequence, not an additional
lensing parameter.  Table~\ref{tab:profiles} lists the three snapshots shown
in Fig.~1 of the Letter; the lensing calculation uses the density profile
itself.

\begin{table}[b]
\caption{Displayed late gravothermal profiles.  Radii are measured in units
of $r_{s,\rm acc}$, {the progenitor's NFW scale radius at host
entry}.  }
\label{tab:profiles}
\begin{ruledtabular}
\begin{tabular}{cccc}
$\log_{10}(\beta\widehat\sigma\widehat t)$ &
$\rho_c/\rho_{s,\rm acc}$ & $r_c/r_{s,\rm acc}$ & $r_{\rm out}/r_{s,\rm acc}$\\
\hline
2.223 & 69.38 & 0.08 & \StageEarlyOuter\\
2.236 & 456.0 & 0.03 & \StageMiddleOuter\\
2.237 & 1178.0 & 0.02 & \StageLateOuter\\
\end{tabular}
\end{ruledtabular}
\end{table}

The isolated  mass profile is normalized to the {progenitor
halo at host entry}.  We then
determine the tidal radius from
\begin{equation}
 4\pi\int_0^\infty r^2\rho_{\rm SIDM}(r;r_t)\,\dd r=M_{\rm sub}
 \label{eq:massmatch}
\end{equation}
for $r_t$.  This removes the outer envelope while preserving the contracted
center.  The fractional mass  uncertainty is below $1.3\times10^{-12}$ for every
representative {model subhalo}.  A global rescaling by
$M_{\rm sub}/M_{\rm prof}$, where $M_{\rm prof}$ is the unstripped profile
mass, would instead dilute the core together with the envelope.  It lowers the
four-copy cross section by 91.7\% [Fig.~\ref{fig:systematics}(a,b)] and is not
used in the event rates.

JVAS B1938+666-$\mathcal V$ provides an observational example of the same
two-scale structure, but it is not used to calibrate our rates.  The favored
lens models place $(1.13\pm0.04)\times10^6\,\Msun$ inside 80 pc and require a
dense unresolved component embedded in an extended envelope
~\cite{2025NatAs...9.1714P,2026NatAs..10..440V}.  Zhang and
Yu~\cite{2026arXiv260612909Z} reproduced this profile by evolving a
$1.5\times10^8\,\Msun$ SIDM progenitor after truncation to a
$2.7\times10^6\,\Msun$ remnant.  The {weighted SASHIMI
population contains model subhalos with comparable accretion-to-lens-epoch
mass loss}.  The measured profile is nevertheless
not inserted into our population: the abundance and collapse weights remain
those of the simulations cited below.  The observation supports the physical
plausibility of a dense core inside a stripped envelope, not the lifetime of
that state.

The truncation in Eq.~\eqref{eq:massmatch} is applied at fixed gravothermal
stage.  A fully coupled tidal-fluid calculation can change the time required
to reach this stage; indeed, the B1938+666 model finds that truncation
accelerates collapse~\cite{2026arXiv260612909Z}.  This affects the transient
weight and the abundance of collapsed objects, but not the lensing threshold
or the cross section of a specified density profile.  The Letter therefore
keeps the transient and long-lived endpoints separate.

\section{Why the second pair has a sharp threshold}

The mean subhalo convergence is
$\bar\kappa_{\rm sub}(<r)=\alpha_{\rm sub}(r)/r$, where $\alpha_{\rm sub}(r)$ is the subhalo deflection angle.  Away from a principal
axis, let $(x_t,x_r)$ and $(y_t,y_r)$ be the image- and source-plane
coordinates along the tangential and radial directions.  The local lens
equation gives
\begin{equation}
 x_t=\frac{y_t}{\lambda_t-\bar\kappa_{\rm sub}(<r)},\qquad
 x_r=\frac{y_r}{\lambda_r-\bar\kappa_{\rm sub}(<r)} .
 \label{eq:coordinates}
\end{equation}
Together with $r^2=x_t^2+x_r^2$, Eq.~\eqref{eq:coordinates} gives Eq.~(2) of
the Letter.  A positive-parity macro image has
$0<\lambda_t<\lambda_r$, with $\lambda_t, \lambda_r$ being eigenvalues of the inverse magnification matrix.  When the maximum mean convergence lies between the
two eigenvalues, only the tangential denominator can vanish and one local
minimum--saddle pair is possible.  Once
$\max_r\bar\kappa_{\rm sub}(<r)>\lambda_r$, the radial denominator can also
vanish, allowing a saddle--maximum pair.  After the SNR and {time-delay} cuts, the
first nonzero four-copy area occurs at
$\max\bar\kappa_{\rm sub}/\lambda_r=\CriticalityOnset$; no configuration at or
below unity yields four detected copies.

For a macro saddle, $\lambda_t<0<\lambda_r$ and
$\bar\kappa_{\rm sub}(<r)>0$.  The tangential denominator cannot vanish, so
the same two-pair sequence is absent.  Saddles can still produce detectable
{triplets, but not the five-image branch of a
minimum for these spherical subhalos.}  Ellipticity can deform the caustics and
shift the threshold, but it does not remove the need for sufficient central
convergence.

\section{Ellipticity and the survival of the five-image branch}

Equation (2) of the Letter is exact for a circular perturber.  To test whether
the second pair is an artifact of that symmetry, we replace the circular
lensing potential $\psi(r)$ by the pseudo-elliptical potential
\begin{equation}
 \psi_q(\bm x)=\psi\!\left(\sqrt{q x_1'^2+x_2'^2/q}\right),
 \label{eq:pseudoelliptical}
\end{equation}
where $(x_1',x_2')$ are rotated by an angle $\varphi$ relative to the macro
shear.  We use $q=1,0.9,0.8$ and
$\varphi=0^\circ,22.5^\circ,45^\circ$, spanning moderate projected
ellipticities and their relative orientations~\cite{1997ApJ...482..604K}.
The full two-dimensional lens equation is solved with the analytic Jacobian of
Eq.~\eqref{eq:pseudoelliptical}; no radial reduction is used.

For $q=1$, the two-dimensional solver returns exactly the roots of the
independent circular solver.  Across all 27 combinations of gravothermal
stage, $q$, and $\varphi$, the largest lens-equation residual is
$4.2\times10^{-12}$ and
$N_{\rm min}+N_{\rm max}-N_{\rm sad}=1$.  Ellipticity does move the caustics:
the source marked in Fig.~1 of the Letter follows $1\to1\to3$ rather than
$1\to3\to5$ for several orientations.  Hence this fixed-source sequence is not
a topological invariant.

The relevant question is whether the five-image branch itself disappears.  We
hold the unperturbed macro-image radius fixed and scan 24 source directions
through the deepest profile.  {Figure~\ref{fig:systematics}(c)} shows that every
tested $(q,\varphi)$ combination retains a five-image region.  The fraction of directions
inside it is 0.75 for the circular lens and ranges from 0.17 to 0.75 after the
deformation.  The second-pair topology is therefore robust, whereas its
source-plane area---and hence the event rate---carries an {ellipticity systematic}.
The main population forecast {uses circular lens models} and does not hide
this uncertainty in a fitted correction factor.

\section{Stationary-image solver}

Equation~\eqref{eq:coordinates} reduces the two-dimensional problem to a
radial root search.  We bracket every root on a logarithmic grid and add finer
grids around each solution of
$\bar\kappa_{\rm sub}=\lambda_t$ or
$\bar\kappa_{\rm sub}=\lambda_r$.  Sources on a principal axis are handled
separately, including the symmetric pair away from that axis.  Adjacent
brackets are merged only when the image-plane separation is below $10^{-7}$
times the larger image radius.

At each root we diagonalize the Hessian of the Fermat potential,
\begin{align}
 \phi(\bm x)&=\tfrac12\bm x^{\mathsf T}{\bm A}\bm x
 -\bm y\!\cdot\!\bm x-\psi(r),\\
 \mu_j&=[\det\nabla\nabla\phi(\bm x_j)]^{-1},
\end{align}
where ${\bm A}=\mathrm{diag}(\lambda_t,\lambda_r)$ is the local macro-lens
matrix and $\psi$ is the subhalo lensing potential.  Two positive Hessian
eigenvalues give a minimum, eigenvalues of different signs give a saddle, and two negative eigenvalues give a maximum.  The
quantity $\mu_j$ is the signed magnification;

We tested 160 random two-dimensional configurations spanning both macro
parities, all three displayed profiles, the subhalo population, and macro
magnifications from 1.2 to 100.  Grids with 1200, 2400, and 4800 radial points
returned the same image number in every case.  The largest relative changes in
arrival time and magnification were $2.0\times10^{-13}$ and
$1.5\times10^{-13}$; the largest lens-equation residual was
$3.4\times10^{-11}$.

\section{Detector band and copy recovery}

{The condition $8\pi G(1+z_l)Mf/c^3\sim1$ places the diffraction transition of $10^6$--$10^{10}\,\Msun$ stripped remnants near $10^{-7}$--$10^{-2}\,\mathrm{Hz}$.}
Here $G$ is Newton's constant, $c$ the speed
of light, $M$ the subhalo mass, $z_l$ the lens redshift, and $f$ the observed
GW frequency.  ET+CE is therefore in the stationary-image limit.  A space-based
detector would probe the continuous diffraction pattern of the same profiles,
which is a different observable from the {time-resolved,
delayed GW copies} considered here.

For a cataloged macro image with network SNR ${\rho_{\rm macro}}$ and signed
magnification ${\mu_{\rm macro}}$, stationary image $j$ has SNR
\begin{equation}
 \rho_j={\rho_{\rm macro}}
 \sqrt{\frac{|\mu_j|}{|{\mu_{\rm macro}}|}} .
\end{equation}
We require $\rho_j\ge8$.  Arrivals separated by less than one second are
merged and represented by the brighter image. {This defines our arrival-clustering rule.}

\begin{figure*}[t]
\includegraphics[width=\textwidth]{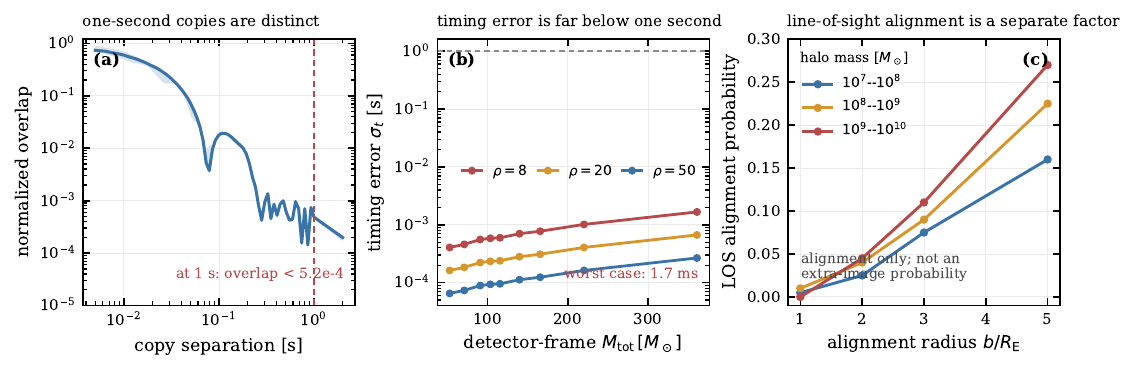}
 \caption{{Panels (a) and (b) test temporal resolvability; panel (c) separately tests the chance alignment of line-of-sight halos.} Detector and line-of-sight checks.  (a) Absolute normalized overlap
 of two identical ET+CE chirps versus their temporal separation.  The band spans nine
 BBH mass quantiles; the vertical line marks the one-second clustering rule.
 (b) Fisher timing precision for the same masses at three network SNRs; the
 gray line marks one second.  (c) Probability that at least one line-of-sight
 halo lies within the indicated number of its Einstein radii of either of two
 macro images.  This is an alignment probability, not a high-order-caustic
probability.}
\label{fig:detector}
\end{figure*}

The arrival-clustering rule interval is much longer than the detector timing resolution.
For nine detector-frame BBH mass quantiles, we generated IMRPhenomXAS
waveforms~\cite{2020PhRvD.102f4001P} and used an effective network inverse
noise spectrum for three ET and two CE detectors.  Figure
~\ref{fig:detector}(a) shows the normalized overlap with a delayed copy.  At
one second the largest overlap is \OneSecondOverlap.  Panel (b) gives the
Fisher timing error, which at network SNR 8 ranges from \TimingLow\ to
\TimingHigh, more than two orders of magnitude below the arrival-clustering rule interval.

\section{Subhalo and strong-lens populations}

We use the public semi-analytical {SASHIMI calculation} for
subhalo accretion and tidal evolution~\cite{2022MNRAS.517.2728H}; the
{SIDM extension of this approach is described in
Ref.~\cite{2025JCAP...02..053A}}.  The fiducial sample has lens redshift $z_l=1$,
source redshift $z_s=5.2$, and host mass $3.4\times10^{12}\,\Msun$.  We
evaluate the subhalo population at the projected radius of each macro image in
the Planck 2018 flat cosmology~\cite{2020A&A...641A...6P}.

For the velocity-dependent SIDM model, we use
$\sigma_0/m_\chi=200\,{\rm cm^2\,g^{-1}}$ and
{$\omega=200\,{\rm km\,s^{-1}}$}.  {The adopted
$\sigma_0$--$\omega$ model is}
\begin{equation}
 \frac{\sigma(v_{\rm rel})}{m_\chi}=
 \frac{\sigma_0/m_\chi}{[1+v_{\rm rel}^2/\omega^2]^2},
 \label{eq:velocity_dependent_cross_section}
\end{equation}
{where $v_{\rm rel}$ is the relative speed of the two scattering SIDM
particles and $\omega$, rather than a halo velocity, is the interaction's
turnover scale}~\cite{2025PhRvD.111f3001Z}.  The collapse probabilities are
taken directly from that reference: 0.347, 0.232, 0.351, and 0.389 in the
accretion-mass bins $10^{8}$--$10^{8.2}$, $10^{8.2}$--$10^{8.5}$,
$10^{8.5}$--$10^9$, and $10^9$--$10^{10}\,\Msun$.  We assign zero
probability below an accretion mass of $10^8\,\Msun$, outside the simulation
calibration.  Here $\sigma_0/m_\chi$ is the low-velocity cross section per
unit dark-matter mass.  {For gravothermal heat transport,
$v_{\rm rel}$ samples particle pairs inside the subhalo; the simulations
characterize its mean by $\langle v_{\rm rel}\rangle\simeq3.8\sigma_v$, with
$\sigma_v$ the one-dimensional internal velocity dispersion.  For
host--subhalo evaporation, the same cross section is instead evaluated at the
bulk orbital relative speed, typically $10^2$--$10^3\,{\rm km\,s^{-1}}$.
These two velocities enter different physical processes.  Because we import
the collapse fractions from simulations containing both tides and
evaporation, we do not insert an additional orbital speed into the internal
heat-transfer mapping.}  We group the {weighted model subhalos} by accretion mass and
redshift,
retained mass fraction, scale radius, scale density, and tidal radius.  Twelve
{representative model subhalos stand for these groups and
carry their population weights}.

The public ET+CE strong-lens catalog represents ten years of observations as provided in
~\cite{2026ApJS..284...56L}.  We expand every BBH into its macro images, retain
minima and saddles, and interpolate the cross-section grid in SNR,
$|{\mu_{\rm macro}}|$, $\lambda_r$, and parity.  Let $R_{\rm E}$ be the
{lens-plane point-mass Einstein radius corresponding to
$M_{\rm acc}$} and $A_{\ge m,ik}$ the source-plane area in which
representative {model subhalo} $i$ splits macro image $k$ into at least $m$ detected
copies.  The corresponding probability is
\begin{equation}
 p_{\ge m,k}=1-\exp\!\left[-\bar N_{\rm sub}(<R_{\rm E})
 \sum_i w_i\frac{A_{\ge m,ik}}{\pi R_{\rm E}^2}\right] .
\end{equation}
Here $w_i$ is the normalized ($\sum_iw_i=1$) population weight and
$\bar N_{\rm sub}(<R_{\rm E})$ is the mean number of eligible subhalos
projected within $R_{\rm E}$.  For CDM halos
selected to collapse, we replace $w_i$ by $w_iP_{{\rm coll},i}$ without
renormalizing, where $P_{{\rm coll},i}$ is the collapse probability above.
{This unnormalized weighting isolates the CDM contribution of exactly the subhalos selected to collapse before those objects are replaced by their evolved SIDM profiles.}

We sample 60 source impact radii from $10^{-3}R_{\rm E}$ to $3R_{\rm E}$,
logarithmically inside $0.3R_{\rm E}$ and linearly outside, together with eight
midpoint angles in one quadrant.  Reflection symmetry supplies the other
quadrants.
Each annulus is weighted by
the fraction of angles that satisfies the copy-number cut.

\section{Collapsed-phase time-average and endpoint replacement}

The deepest collapsed-phase snapshot is not itself a population model.  For the
transient case, we begin when the central density reaches five times its
initial value, the collapse definition used in the population simulations~\cite{2025PhRvD.111f3001Z}.
This occurs at $\log_{10}(\beta\widehat\sigma\widehat t)=2.00$. We
integrate uniformly in the $\beta\widehat\sigma\widehat t$ variable to
the formal-collapse value $\log_{10}(\beta\widehat\sigma\widehat t)=2.238$ with
eight-point Gauss--Legendre quadrature.  The last node has
$\rho_c/\rho_{s,\rm acc}=312$ but carries only 5.06\% of the post-threshold {collapsed-phase} time weight.  The deepest displayed profile,
$\rho_c/\rho_{s,\rm acc}=1178$, lies still closer to formal collapse and is
used only for the long-lived-remnant endpoint.

For $m=3,4,5$, the expected count of BBH macro-image candidates is
\begin{equation}
 N_{\ge m}=N_{\ge m}^{\rm CDM}
 -N_{\ge m}^{\rm CDM,coll}+N_{\ge m}^{\rm SIDM} .
 \label{eq:replacement}
\end{equation}
The second term removes CDM halos selected to collapse; the third inserts their
evolved SIDM profiles.  Table~\ref{tab:rates} gives the BBH results.  The
transient phase barely changes the triplet count and rarely retains four or
five detectable copies.  A long-lived compact remnant gives a much larger
high-multiplicity fraction.  These are counts of macro-image candidates, not
counts of distinct astrophysical mergers; different macro images of one
strong-lens system are separate local lens environments in this calculation.

\begin{table}[t]
\caption{Expected BBH macro-image candidates in five years of ET+CE operation.  The transient and
remnant rows include the CDM subtraction in Eq.~\eqref{eq:replacement}. }
\label{tab:rates}
\begin{ruledtabular}
\begin{tabular}{lccc}
population & $N_{\ge3}$ & $N_{\ge4}$ & $N_{\ge5}$\\
\hline
CDM & 0.36 & $1.55\times10^{-6}$ & $8.63\times10^{-9}$\\
transient collapse & 0.50 & $2.24\times10^{-3}$ & $1.98\times10^{-3}$\\
long-lived remnant & 1.66 & 0.15 & 0.08\\
\end{tabular}
\end{ruledtabular}
\end{table}

Dividing the last two columns by $N_{\geq3}$ gives the global high-copy
fractions listed below.  Figure~3(b) of the Letter gives a complementary view:
it bins the candidate counts by total macro-lens magnification and hatches the
contribution with at least four copies.  The four-copy fractions are
\CDMFourFraction, \TransientFourFraction, and \RemnantFourFraction\ for CDM,
transient collapse, and a long-lived remnant; the corresponding five-copy
fractions are \CDMFiveFraction, \TransientFiveFraction, and
\RemnantFiveFraction.

\begin{figure*}[t]
\includegraphics[width=\textwidth]{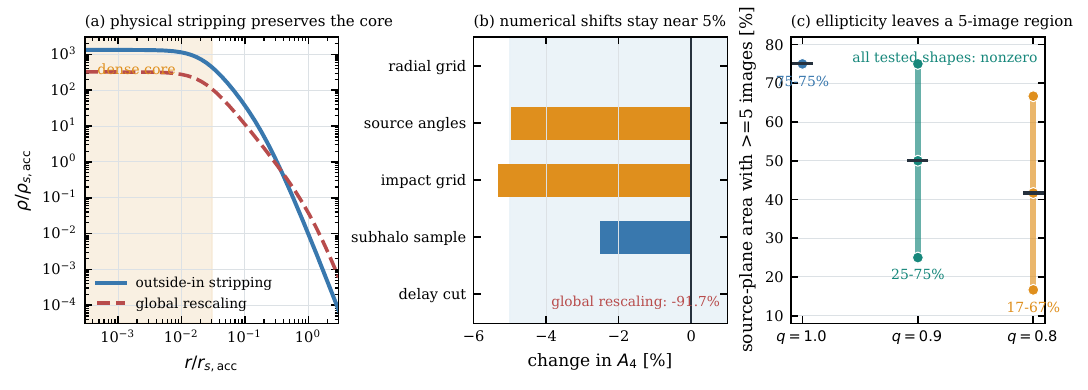}
 \caption{{Compact robustness summary.  (a) Outside-in tidal stripping
 preserves the collapsed center, unlike an unphysical global density
 rescaling.  (b) Resolution and sampling changes shift the four-copy cross
 section only at the few-percent level; global rescaling is shown separately
 because it changes the physical profile rather than the numerical
 resolution.  (c) Across the tested projected axis ratios and orientations,
 the deep-collapse profile retains a nonzero five-image source-plane area.}}
\label{fig:systematics}
\label{fig:ellipticity}
\end{figure*}

\section{Compound CDM and line-of-sight structure}

The single-halo CDM row in Table~\ref{tab:rates} assumes the fiducial
concentration attached to each {weighted SASHIMI model
subhalo}.  We first repeat the complete
calculation after multiplying every {lens-epoch concentration}
at fixed {lens-epoch bound mass} by three or six.  Nothing else
is changed: the subhalo weights, ET+CE
macro-image catalog, parity and eigenvalue interpolation, source positions,
SNR threshold, and one-second {time-delay} cut are identical.  Table
~\ref{tab:same_selection_cdm} is therefore directly comparable to the SIDM
rows in Table~\ref{tab:rates}.  A factor of three raises the triplet count but
leaves the four-copy count thirteen times below the long-lived-remnant result.
The extreme factor-six model reaches a comparable high-copy fraction and is a
genuine degeneracy.  These multipliers bracket a stress test; the fiducial
concentration--mass relation and its physical scatter remain the baseline
~\cite{2019ApJ...871..168D}.

\begin{table}[t]
\caption{Same-selection single-halo CDM counts for BBH macro images in five
years of ET+CE operation.  Concentration is multiplied at fixed {lens-epoch bound mass}.  }
\label{tab:same_selection_cdm}
\begin{ruledtabular}
\begin{tabular}{lccc}
CDM concentration & $N_{\ge3}$ & $N_{\ge4}$ & $N_{\ge5}$\\
\hline
$c_{\rm fid}$ & 0.36 & $1.55\times10^{-6}$ & $8.63\times10^{-9}$\\
$3c_{\rm fid}$ & 1.45 & 0.01 & 0.01\\
$6c_{\rm fid}$ & 3.18 & 0.23 & 0.11\\
\end{tabular}
\end{ruledtabular}
\end{table}

The single-halo calculation is still not the full null hypothesis.  Vujeva
et al.~\cite{2025arXiv251014953V} simulated subhalo
populations near highly magnified images.  Their probability for more than
five highly magnified images is conditional on a smooth-lens total
magnification above 25.  Applying that probability to all 617 lensed systems
per year would overstate the background.  The same magnification cut, together
with at least four macro images, selects 66 BBH systems in the ten-years public
catalog, or 33 in the {five-year catalog}.

Table~\ref{tab:compound} multiplies this matched count by each probability in
their Table I.  The lower value uses the subhalo-only calculation; the upper
value doubles it as a conservative allowance for the omitted line-of-sight halos.
This factor is not a substitute for a multi-plane calculation.  As an
independent check, we generated line-of-sight halo populations and counted
halos within one to five Einstein radii of either of two macro images.
Figure~\ref{fig:detector}(c) shows the resulting alignment probability.  It
supports an order-unity allowance but does not predict image multiplicity.

\begin{table}[t]
\caption{Expected compound-CDM counts for the selected ET+CE systems.
$P_{>5}$ is the published conditional probability; the last column runs from
the subhalo-only result to twice that value as an allowance for line-of-sight
halos. }
\label{tab:compound}
\begin{ruledtabular}
\begin{tabular}{lcc}
CDM model & $P_{>5}$ & candidates in 5 yr\\
\hline
fiducial & $2.8\times10^{-4}$ & 0.01--0.02\\
$3c$ & $6.4\times10^{-3}$ & 0.21--0.42\\
$6c$ & $1.6\times10^{-2}$ & 0.53--1.06\\
$\Sigma_{\rm sub}=0.05\,{\rm kpc^{-2}}$ & $1.7\times10^{-4}$ & 0.01--0.01\\
$\Sigma_{\rm sub}=0.07\,{\rm kpc^{-2}}$ & $3.1\times10^{-4}$ & 0.01--0.02\\
\end{tabular}
\end{ruledtabular}
\end{table}

In the last two rows of Table~\ref{tab:compound}, $\Sigma_{\rm sub}$ denotes
the projected surface density of eligible subhalos.  The published
high-multiplicity definition differs from the local four-copy
selection used here.  Table~\ref{tab:compound} is therefore a comparison, not
a background subtraction.  Data analysis must simulate the compound
population and fit the copy number, parity, relative delays, relative
magnifications, and macro lens jointly.  The present comparison already shows
that concentration matters much more than subhalo abundance.

\begin{figure*}[t]
\includegraphics[width=\textwidth]{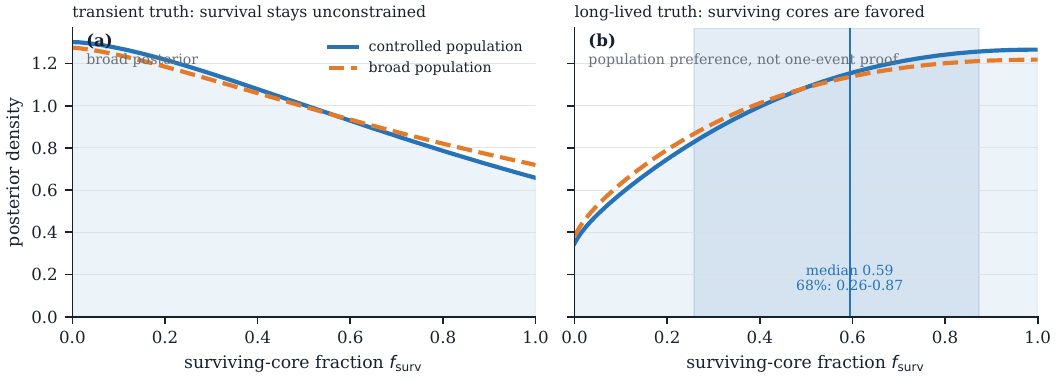}
 \caption{{The two colors use the same likelihood and differ only in the allowed CDM-concentration range.} Ten-year forecast posterior for the surviving collapsed-core
 fraction.  (a) Asimov data generated at the transient endpoint.  (b) Asimov
 data generated at the long-lived endpoint.  Blue line restricts the CDM
 concentration to $c/c_{\rm fid}\le3$; red allows the extreme factor-six stress
 test.  Both curves marginalize a 50\% log-rate calibration.  The broad
 posteriors reflect the small absolute event sample and are part of the forecast,
not hidden by quoting only the endpoint rate ratio.}
\label{fig:survival_posterior}
\end{figure*}

\section{Likelihood for the surviving collapsed-core fraction}

The Letter introduces $f_{\rm surv}$ as the fraction of collapse-selected
halos that retain the deepest calibrated profile.  For each inclusive
copy-number threshold, its rate is
\begin{equation}
 N_{\ge m}(f_{\rm surv})=(1-f_{\rm surv})N_{\ge m}^{\rm tr}
 +f_{\rm surv}N_{\ge m}^{\rm rem}.
\end{equation}
We use two disjoint counting bins,
$\Lambda_3=N_{\ge3}-N_{\ge4}$ and $\Lambda_4=N_{\ge4}$, so that no candidate is
counted twice.  {We conservatively add the full same-selection CDM concentration shift to both inclusive rates, including the noncollapsed component.}
\begin{equation}
 \bm N(f_{\rm surv},c)=\bm N(f_{\rm surv},c_{\rm fid})
 +\bm N_{\rm CDM}(c)-\bm N_{\rm CDM}(c_{\rm fid}).
 \label{eq:concentration_nuisance}
\end{equation}
Here $\bm N=(N_{\ge3},N_{\ge4})$ is the vector of inclusive candidate rates.
The rates are interpolated in $\ln c$ and
$\ln N$ between the three calculated concentrations in Table
~\ref{tab:same_selection_cdm}.  Equation~\eqref{eq:concentration_nuisance} deliberately gives the
concentration uncertainty for its full CDM change even though only the
noncollapsed part of the population should receive it; the resulting
survival constraint is conservative.

For an exposure time $T$, the count likelihood is
\begin{equation}
 {\cal L}=\prod_{b=3,4}{\rm Poisson}
 \left[n_b\mid a(T/5\,{\rm yr})\Lambda_b(f_{\rm surv},c)\right],
\end{equation}
where $n_b$ is the observed count in bin $b$ and
$b=3,4$ labels the two disjoint copy-number bins.
We assign to the common rate calibration $a$ a lognormal prior
centered at unity (median one) with width $\sigma_{\ln a}=0.5$ in $\ln a$.
This single nuisance moves the
two counts together and therefore preserves their ratio.  Here $c$ is the CDM
concentration and $c_{\rm fid}$ its baseline value.  The controlled
 concentration prior is flat in $\ln c$ over $c/c_{\rm fid}\in[1,3]$;
{the stress-test prior extends the upper limit to $6c_{\rm fid}$.}  The prior on $f_{\rm surv}$ is uniform on
$[0,1]$.

Figure~\ref{fig:survival_posterior} shows the ten-year Asimov forecasts.
If the
long-lived endpoint is true at the fiducial concentration, the controlled
calculation gives
$f_{\rm surv}=\SurvivingPosteriorMedian^{+0.29}_{-0.34}$ (central 68\%).  The
interval is broad because the expected high-copy count is only 0.30 in ten
years of ET+CE operation.  Extending the concentration prior to $6c_{\rm fid}$ shifts the median
to 0.54 and the interval to $[0.20,0.85]$.  These numbers are not a promised
detection significance.  They state the exposure-limited information in the
forecast and show why electromagnetic concentration constraints belong in the
measurement.  {Such lensing or stellar-dynamical information can enter a future joint analysis as an external prior on $c$; no such external constraint is assumed in the rates quoted here.}

\section{Which control observables actually reduce the risk?}

\begin{figure*}[t]
\includegraphics[width=\textwidth]{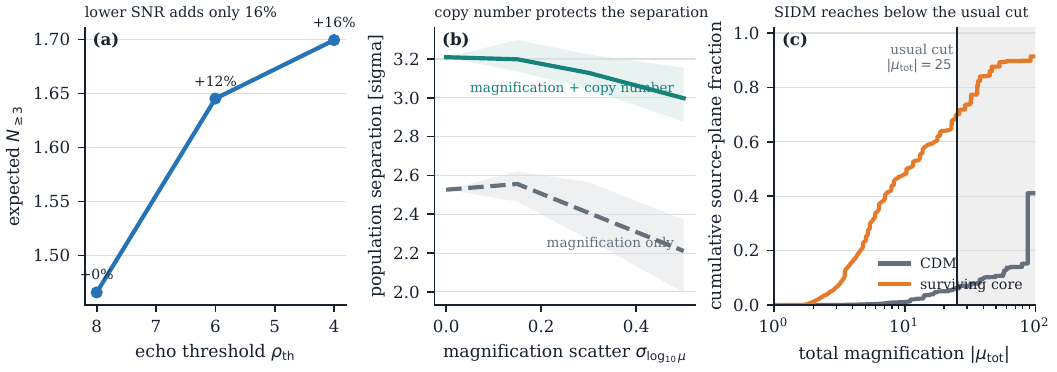}
 \caption{{The panels progress from search sensitivity, to discriminating morphology, to the macro-magnification selection that suppresses false positives.} Quantitative risk-control tests.  (a) Three-copy yield in one common
 population sample as the faint-copy SNR threshold is lowered; the gain
 saturates.  (b) Shape-only Asimov separation from single-halo CDM using system
 magnification (blue) and system magnification plus copy number (red).  Shading
shows the central 68\% over realizations of the stated magnification uncertainty;
the dashed line uses copy number alone.  (c) Weighted total-magnification CDF
of detected candidates.  The gray region marks the selection used by the
published compound-CDM calculation.}
\label{fig:risk_control}
\end{figure*}

We tested three possible remedies using the same candidate-level population
sample.  The first is a targeted search for faint copies after a loud arrival
has fixed the detector-frame masses, spins, and sky response.  In a matched
sample, lowering the faint-copy threshold from 8 to 6 raises the three-copy
yield by 12\%; lowering it to 4 raises the yield by 16\%
[Fig.~\ref{fig:risk_control}(a)].  Most candidates already have three copies
above threshold, so targeted recovery does not change the rate-limited nature
of the experiment.  We therefore do not credit it in the Letter's rates.

The second test uses the total magnification of the smooth macro system, the
quantity on which the published compound-CDM calculation is selected.  Among
detected candidates, its weighted median is $|\mu_{\rm tot}|=10.3$ for the
collapsed-core population and $|\mu_{\rm tot}|=194$ %194
for the single-halo CDM control.  Only 30.1\% of the former, compared with
93.8\% of the latter, lie above $|\mu_{\rm tot}|=25$.  After
normalizing each distribution to the rates given in the Letter, surviving cores
give 1.16 candidates below 25 in five years, whereas standard CDM gives
0.022.  An ordinary CDM halo therefore {requires a smaller
local macro-lens eigenvalue $\lambda_r$} to cross the pair-creation threshold.

We quantify only the shape information, profiling out the common rate.  Here
$N_{\rm copy}$ denotes the number of detected copies assigned to one
macro-image candidate.  The quoted Asimov separation is the square root of the
likelihood-ratio test
statistic after profiling over that common rate.  %Copy number
The number of detected copies alone gives an
Asimov separation of 2.18 in the five-year matched
sample, while the joint $(|\mu_{\rm tot}|,N_{\rm copy})$ distribution gives
3.21.  Host identification, the macro model, and the mass-sheet degeneracy
broaden the inferred magnification.  We therefore perturb each value in
$\log_{10}|\mu_{\rm tot}|$.  With 0.30 dex scatter the median joint separation
is 3.13; even with 0.50 dex it is 3.00
[Fig.~\ref{fig:risk_control}(b)].  These numbers compare the collapsed-core
sample with the calculated single-halo CDM control; they are not a claimed
significance against every compound or baryonic lens.  Their role is to show
that the lens environment carries information independent of the event count.

We then test the lower-magnification window against the line-of-sight structure
directly.  One hundred independent pyHalo CDM light cones are applied to eight
common strongly lensed sources with $|\mu_{\rm tot}|<25$.  For every one of
the resulting 800 configurations, we solve the full multi-plane lens equation.
We retain all field halos above $10^{8.5}\,\Msun$ and lower-mass halos within
$0.3''$ of a smooth macro ray.  A false local pair must contain two additional
images with $|\mu|>1$, separated by less than $0.05''$, whose midpoint lies
within two Einstein radii of the same field halo.  None is found, giving a
one-sided 95\% binomial upper limit of 0.37\% per tested configuration.  In 11
cases the number of macro images changes, but the added images are not
localized on a field halo.  These are the changes to the macro caustic, which a
joint macro model must fit, rather than compact-perturber copies.

The third test is a halo shape.  Averaging uniformly over the nine calculated
combinations of $q=0.8,0.9,1$ and orientations reduces the angular coverage of
the five-image branch to 0.74 of the circular model value.  This is a diagnostic,
not a population correction, because a physical $p(q)$, the probability
distribution of projected axis ratios, and a full source-plane
integral are still required.  It shows that ellipticity preserves the branch
but cannot rescue the rate.  Accordingly, the main conclusions are: targeted
recovery helps validate a candidate, the joint magnification--multiplicity
distribution controls false positives, and nonsphericity must enter as a
downward rate systematic.

\section{Numerical and population checks}

{Figure~\ref{fig:systematics} collects the three robustness checks most
directly tied to the inference.  Outside-in stripping preserves the collapsed
center, the source integration is stable at the 5\% level, and every tested
ellipticity retains a nonzero five-image region.  The parity and redshift
checks remain documented in the surrounding text and numerical tables rather
than being repeated as separate visual panels.  Table~\ref{tab:convergence}
lists the individual convergence tests.}

\begin{table}[t]
\caption{Changes in the mean four-copy area relative to the primary
calculation.}
\label{tab:convergence}
\begin{ruledtabular}
\begin{tabular}{lcc}
test & change & $\Delta A_4/A_4$\\
\hline
radial brackets & $1200\rightarrow2400$ & \ConvRadial\\
source angles & $8\rightarrow12$ & \ConvAngle\\
impact radii & $60\rightarrow80$ & \ConvImpact\\
representative {model subhalos} & $12\rightarrow20$ & \ConvMedoid\\
arrival clustering & $1\rightarrow2\,{\rm s}$ & \ConvDelay\\
\end{tabular}
\end{ruledtabular}
\end{table}

The dominant remaining uncertainties are physical: the abundance of
black-hole-seeded remnants, line-of-sight populations outside the explicit
light-cone audit~\cite{2019MNRAS.487.5721G}, the shape distribution of deeply
collapsed subhalos, the redshift dependence of the collapse fraction,
baryonic satellites, and the survival time of the collapsed core.
The Letter keeps the last uncertainty explicit by quoting transient and
long-lived endpoints separately.

\end{document}